\documentclass[bimj,fleqn]{w-art}
\usepackage{times}
\usepackage{w-thm}
\usepackage[authoryear]{natbib}
\usepackage{amsfonts, amsthm, amssymb,amsmath}
\usepackage{adjustbox}      
\usepackage{bm, bbm, booktabs}
\usepackage{color}
\usepackage{comment}
\usepackage{epstopdf, epsfig}
\usepackage{etoolbox}       
\usepackage{graphicx}
\usepackage{hyperref}       
\usepackage{natbib}
\usepackage{mathrsfs}
\usepackage{multirow}
\usepackage{threeparttable} 
\usepackage{setspace}       
\usepackage[linesnumbered, ruled, vlined]{algorithm2e}
\usepackage{siunitx}
\usepackage{adjustbox}      
\usepackage{threeparttable}
\usepackage{comment}
\usepackage[normalem]{ulem}

\newtheorem{mytheorem}{Theorem}[section]

\newtheorem{myexample}[mytheorem]{Example}

\makeatletter
\newcommand{\algorithmfootnote}[2][\footnotesize]{%
  \let\old@algocf@finish\@algocf@finish
  \def\@algocf@finish{\old@algocf@finish
    \leavevmode\rlap{\begin{minipage}{\linewidth}
    #1#2
    \end{minipage}}%
  }%
}
\makeatother

\SetCommentSty{mycommfont}

\graphicspath{{figures/}}

\theoremstyle{plain}

\theoremstyle{definition}

\usepackage[]{graphicx}
\chardef\bslash=`\\ 

\begin{document}
\DOIsuffix{bimj.200100000}
\Volume{52}
\Issue{61}
\Year{2010}
\pagespan{1}{}
\keywords{weighted parametric testing; group sequential design; sequential p-values; multiplicity; adjusted p-value\\
\noindent \hspace*{-4pc} \\
\noindent\hspace*{-4.2pc} Supporting information for this article is available at \url{https://merck.github.io/wpgsd}. 
}  

\title[Adjusted inference for multiplicity in GSD]{Adjusted inference for multiple testing procedure in group sequential designs}
\author[Yujie Zhao {\it{et al.}}]{Yujie Zhao\footnote{Corresponding author: {\sf{e-mail: yujie.zhao@merck.com}}}\inst{,1}} 
\address[\inst{1}]{Merck \& Co., Inc., Rahway, NJ, USA}
\author[dd]{Qi Liu\inst{1}}
\author[]{Linda Z. Sun\inst{1}}
\author[]{Keaven M. Anderson\inst{1}}
\Receiveddate{2023/08} \Reviseddate{TBD} \Accepteddate{TBD} 

\begin{abstract}
Adjustment of statistical significance levels for repeated analysis  in group sequential trials has been understood for some time.
Similarly, methods for adjustment accounting for testing multiple hypotheses are common.
There is limited research on simultaneously adjusting for both multiple hypothesis testing and multiple analyses of one or more hypotheses. We address this gap by proposing \textit{adjusted-sequential p-values} that reject an elementary hypothesis when its adjusted-sequential p-values are less than or equal to the family-wise Type I error rate (FWER) in a group sequential design. 
We also propose sequential p-values for intersection hypotheses as a tool to compute adjusted sequential p-values for elementary hypotheses.
We demonstrate the application using weighted Bonferroni tests and weighted parametric tests, comparing adjusted sequential p-values to a desired FWER for inference on each elementary hypothesis tested.
\end{abstract}

\maketitle                   






\section{Introduction}
\label{sec: introd}

Recent decades have witnessed an increasing trend to answer multiple clinical questions within a single trial while strictly controlling Type I error for all statistical testing performed. Clinical questions may be related to hypotheses concerning multiple populations; in oncology, this can particularly be related to disease type or biomarkers that may be predictive of treatment effectiveness. Trials may also assess multiple experimental arms versus a common control. These efforts result in increasing complexity to adjust for \textit{multiplicity} in the statistical framework. In addition, modern clinical trials are often designed in an adaptive way to allow using data accumulated in the trial to modify the trial's course. Group sequential design (GSD) has been one of the most widely used adaptive designs, in which multiple analyses (i.e., interim and final analyses) may be conducted according to pre-determined rules to enable decisions. Existing literature on multiplicity and GSD focuses on either adjusting multiple hypotheses in a fixed design or adjusting multiple analyses in GSD with a single hypothesis. There is limited research on simultaneously adjusting for both multiple hypotheses and multiple analyses. We address this gap in this paper.

There are two approaches to adjust for multiplicity. One approach is to adjust the significance levels and compare the nominal p-values with corresponding adjusted significance levels. The other approach is to adjust the p-values and compare the adjusted p-values with the overall FWER(e.g., one-sided 0.025). For simple multiple testing procedures, these two approaches are often exchangeable. For example, to test $m$ equally weighted hypotheses using the Bonferroni procedure at the FWER $\alpha$, the adjusted significance level is $\alpha/m$ for each hypothesis and the adjusted p-value for hypothesis $H_i$ is $\min(1, mp_i)$, where $p_i$ is the nominal p-value of hypothesis $H_i$. For more complicated multiple testing procedures, the adjusted p-value approach may be viewed as more straightforward and less confusing since we only need to compare the adjusted p-value of each individual hypothesis with a common benchmark, the FWER for the testing procedure. In this paper, we focus on the adjusted p-value approach. Below is a brief literature review on application of adjusted p-values for multiple testing in clinical trial settings.

For trials with a fixed design, the p-value of a single hypothesis is defined as the probability of obtaining a test statistic at least as extreme as observed under the null hypothesis. This is often called the unadjusted p-value or the nominal p-value. To test multiple hypotheses simultaneously, p-values need to be adjusted according to multiple test procedures to control FWER. \cite{westfall1993res} provide a general definition of the adjusted p-value as the smallest significance level at which one would reject the hypothesis using the given multiple testing procedure. As illustrated above for the Bonferroni procedure, for most simple non-parametric or semi-parametric approaches, the calculation of multiplicity-adjusted p-values is fairly straightforward. For more complicated multiple testing procedures where the closed principle is followed (i.e., a hypothesis $H_i$ can be rejected if all intersection hypotheses containing $H_i$ are rejected \citep{Marcus1976}), this definition can be applied as follows: assuming $p_J$, $J\subseteq\{1, ..., m\}$, is the nominal p-value for intersection hypotheses $H_J$, the adjusted p-value for hypothesis $H_i$ is the largest p-value associated with the index set including $i$, i.e., $\max_{J:i\in J} p_J$ \citep{dmitrienko2019}. \cite{xi2017unified} applied this definition and provided a unified framework for a weighted parametric multiple testing procedure, in which the adjusted p-values can be calculated based on an analytic formula to avoid numerical root finding under multidimensional integration.

For trials with GSD, testing hypotheses at multiple analyses and the correlation among test statistics of interim analyses and final analyses adds a layer of complexity. For a single hypothesis with GSD, \cite{liu2008adaptive} defined the sequential p-value as the minimum significance level at which the hypothesis can be rejected at or before a given analysis time; this can be interpreted as the p-value adjusted for multiple analyses. For multiple hypotheses with GSD, \cite{anderson2022unified} extended the framework of \cite{xi2017unified} by considering the correlation structure among test statistics at interim and final analyses. However, their algorithm focused on the calculation of adjusted boundaries since no analytic formula can be easily provided for adjusted p-values as in \cite{xi2017unified}.

In this paper, we propose the \textit{adjusted-sequential p-value}, which falls in the framework of the adjusted p-value approach. It is a p-value for each elementary hypothesis adjusted for both testing multiple hypotheses and multiple analyses. The rest of the paper is organized as follows. In Section \ref{sec: 4 type of p-values}, we articulate the definition of the proposed adjusted-sequential p-values. In Section \ref{sec: adj-seq p in wpgsd and bonferroni}, we present the implementation of adjusted-sequential p-values in both weighted Bonferroni and weighted parametric tests. Section \ref{sec: discussion} provides conclusions and discussion.

\section{Multiple comparisons using p-values}
\label{sec: 4 type of p-values}

Multiple hypotheses in clinical trials can be due to multiple treatment comparisons (e.g., multiple experimental treatment arms versus a common control), multiple populations (e.g., biomarker-positive versus an overall population of patients), and different endpoints (e.g., progression free survival [PFS] and overall survival [OS]). 
For notation, we denote the \textit{elementary hypotheses} by $H_1, H_2, \ldots, H_m$ for some $m > 0$. 
We also consider \textit{intersection hypothesis} incorporating any non-empty subset of elementary hypotheses. 
That is, for any $\emptyset \neq J \subseteq I = \{1, 2, \ldots, m\}$, we define $H_J = \cap_{j \in J} H_j.$
Controlling the family-wise Type I error $\alpha$ for all elementary hypotheses requires a closed-testing procedure evaluating tests of all $2^m - 1$ intersection hypotheses \citep{Marcus1976}. 

We investigate the above elementary/intersection hypotheses under group sequential designs with a total of $K$ analyses (including the final analysis).
To test the $m$ elementary hypotheses at the $K$ analyses, we utilize the standardized multivariate normal test statistic as $Z_{i, k}$ with $i \in I$ and $k = 1, 2, \ldots, K$ as described in further detail in appendix D and \cite{anderson2022unified}. 
It is not necessary for all hypotheses to be evaluated at all analyses, but we assume a common number of analyses for each hypothesis here to simplify notation and presentation.

With the $m K$ test statistics $\{Z_{i,k}\}_{i \in I, k = 1, \ldots, K}$, we can accept or reject the null hypothesis based on p-values. In multiple testing under group sequential designs, there are commonly four types of p-values.

\textbf{Nominal p-value}. The probability under the null hypothesis that the test statistic of an elementary hypothesis is at least as extreme as observed.

\textbf{Repeated p-value}. For an elementary hypothesis $H_j$ at analysis $k$, the repeated p-value is the minimum significance level that $H_j$ can be rejected at analysis $k$ \citep{JTBook}:
\begin{eqnarray}
\label{equ: def of rep p of H_j}
  p_{j, k}^{rep}
  & = & 
  \inf 
  \left\{ 
    \mu: 
    Z_{j,k} \geq \widetilde Z_{j, k}(\mu) 
  \right\},
\end{eqnarray}
where the boundary $\widetilde Z_{j, k}(\mu)$ is a function of $\mu\in (0,1)$ . Note that the repeated p-values require smoothness assumptions for group sequential bounds to enable the Bonferroni-based method of 
\cite{maurer2013multiple}.

\textbf{Sequential p-value}. For an elementary hypothesis $H_j$ at analysis $k$, the sequential p-value is the minimum significance level that $H_j$ can be rejected at or before analysis $k$:
\begin{footnotesize}
\begin{eqnarray}
\label{equ: def of seq p of H_j}
  p_{j,k}^{seq}
  & = & 
  \inf 
  \left\{ 
    \mu: 
    \underbrace{
    \left\{ 
      Z_{j,1} \geq  \widetilde Z_{j,1}(\mu)
    \right\}
    }_{\text{1st analysis}}
    \; \cup \; 
    \underbrace{
    \left\{ 
      Z_{j,2} \geq  \widetilde Z_{j,2}(\mu) 
    \right\}
    }_{\text{2nd analysis}}
    \;  \cup \; 
    \ldots
    \;  \cup \; 
    \underbrace{
    \left\{ 
      Z_{j,k} \geq  \widetilde Z_{j,k}(\mu) 
    \right\}
    }_{k\text{-th analysis}}
  \right\}.
\end{eqnarray}
\end{footnotesize}
To ease computation, one can simplify the above definition as shown in Appendix \ref{appendix: seq-p simplify}. In addition to the sequential p-value for an elementary hypothesis $H_{j}$, one can also get the sequential p-value for an intersection hypothesis $H_J$ as
\begin{footnotesize}
\begin{eqnarray}
\label{equ: def of seq p of H_J}
  p_{J,k}^{seq}
  & = & \nonumber
  \inf 
  \left\{ 
  \mu: 
  \underbrace{
  \left\{ 
    Z_{j_1,1} \geq  \widetilde Z_{j_1,1}(\mu, J)
  \right\}
  \; \cup \; 
  \left\{ 
    Z_{j_2,1} \geq  \widetilde Z_{j_2,1}(\mu, J) 
  \right\}
  \;  \cup \; 
  \ldots
  \;  \cup \; 
  \left\{ 
    Z_{j_L,1} \geq  \widetilde Z_{j_L,1}(\mu, J) 
  \right\} 
  }_{\text{1st analysis}}
  \right. \\
  & & 
  \; \cup \;
  \underbrace{
  \left\{ 
    Z_{j_1,2} \geq  \widetilde Z_{j_1,2}(\mu, J)
  \right\}
  \; \cup \; 
  \left\{ 
    Z_{j_2,2} \geq  \widetilde Z_{j_2,2}(\mu, J) 
  \right\}
  \;  \cup \; 
  \ldots
  \;  \cup \; 
  \left\{ 
    Z_{j_L,2} \geq  \widetilde Z_{j_L,2}(\mu, J) 
  \right\}
  }_{\text{2nd analysis}}\\
  & & \nonumber
  \left. 
  \; \cup \; 
  \ldots 
  \cup
  \underbrace{
  \left\{ 
    Z_{j_1,k} \geq  \widetilde Z_{j_1,k}(\mu, J)
  \right\}
  \; \cup \; 
  \left\{ 
    Z_{j_2,k} \geq  \widetilde Z_{j_2,k}(\mu, J) 
  \right\}
  \;  \cup \; 
  \ldots
  \;  \cup \; 
  \left\{ 
    Z_{j_L,k} \geq  \widetilde Z_{j_L,k}(\mu, J) 
  \right\}
  }_{k\text{-th analysis}}
  \right\},
\end{eqnarray}
\end{footnotesize}
where $J = \{j_1, j_2, \ldots, j_L\}$ and $\widetilde Z_{j, k}(\mu, J)$ is the upper bound of the elementary hypothesis $H_j$ at the $k$-th analysis, which is decided by $\mu$ and the intersection set $J$.

To calculate the sequential p-value, we first compute the  boundaries $\widetilde Z_{j,k}(\mu, J)$, which depends on the selection of the multiple test procedure.
For example, weighted Bonferroni and WPGSD (see Section \ref{sec: adj-seq p in wpgsd and bonferroni}) are two procedures of the graphical approach (see a review in Appendix \ref{appendix: graphic approach})
With the boundaries available, we use a root finding method to determine $\mu$ in \eqref{equ: def of seq p of H_j} and \eqref{equ: def of seq p of H_J}.

\textbf{Adjusted sequential p-value}. For an elementary hypothesis $H_j$ at analysis $k$, the adjusted-sequential p-value is the maximum of the sequential p-values of all the intersection hypotheses containing $H_j$:
\begin{equation}
\label{equ: def of adj-seq p of H_j}
  p_{j,k}^{aseq} 
  = 
  \max_{\{J: \; j \in J\}} p_{J,k}^{seq}.
\end{equation}
The adjusted-sequential p-value of $H_j$ can be compared with the FWER (e.g., 0.025 one-sided); i.e., if the adjusted sequential p-value of $H_j$ is less than the FWER, then $H_j$ can be rejected. 

One strength of the adjusted-sequential p-value is its simplicity: users only need to compare an adjusted-sequential p-value for any individual hypothesis with the pre-specified FWER (say, 0.025).
In this way, those interpreting the trial are no longer required to utilize many Z boundaries.
In contrast, repeated p-values have varied benchmarks requiring users to compare with allocated alpha, which depends on other test results. In practice, users may not have comprehensive knowledge of the assigned $\alpha$-values over time, making the repeated p-value implementation challenging.

\section{Implementation of adjusted-sequential p-values}
\label{sec: adj-seq p in wpgsd and bonferroni}
The steps for computing the adjusted-sequential p-values are as outlined below.

\begin{itemize}
    \item \textbf{Step 1:} Select the testing approach. Two approaches are discussed here.
    \begin{itemize}
      \item The weighted Bonferroni test is simply illustrated, communicated, and implemented by the graphical approach \citep{bretz2009graphical, burman2009recycling}. For each elementary hypothesis $H_j$, we  reject it if $p_j \leq w_j(I) \alpha$, where $p_j$ is the hyopothesis' nominal p-value. When an elementary hypothesis $H_j$ is rejected, the graph and transition matrix is updated. 
      Testing then proceeds with the reduced hypothesis set excluding $H_j$ and the updated weights. This reject, reallocate and reduce procedure continues until no remaining elementary hypothesis can be rejected. A comprehensive review is given in \cite{bretz2009graphical} and \cite{maurer2013multiple}. 
      \item WPGSD (weighted parametric group sequential design) test. Unlike the weighted Bonferroni test, the WPGSD takes into account the correlations between test statistics when evaluating multiple hypotheses in group sequential designs. This enables less stringent bounds while ensuring robust control of the FWER at the specified level $\alpha$. For a detailed examination, refer to the comprehensive review provided in  \citep{anderson2022unified}.
    \end{itemize}

    \item \textbf{Step 2:} For analysis $k$, compute the sequential p-values through analyis $k$ of all possible intersection hypotheses using the formula in \eqref{equ: def of seq p of H_J}. We consider the interaction hypothesis $H_J$ as an example. Assuming a specified family-wise significance level $\mu$, we can compute the Z boundaries for each elementary hypothesis within $J$ by utilizing the spending function, weights, and chosen testing procedure (weighted Bonferroni or WPGSD). By comparing these Z boundaries with the observed Z statistics, we can reject $H_J$ if any observed Z statistic exceeds the respective Z boundary. We employ root-finding to identify the minimum significance level $\mu$ at which $H_J$ can be rejected. This identified root corresponds to the sequential p-value of $H_J$ at analysis $k$, as defined in \eqref{equ: def of seq p of H_J}.

    \item \textbf{Step 3:} Compute the adjusted sequential p-value for each elementary hypothesis $H_j$ by taking the maximum of all the sequential p-value of the intersection hypotheses containing $H_j$, as shown in \eqref{equ: def of adj-seq p of H_j}.

    \item \textbf{Step 4:} Repeat Steps 2 and 3 for all analyses. 

\end{itemize}

In Step 2, the procedure of computing Z boundaries is outlined in \cite{maurer2013multiple} when the weighted Bonferroni method is chosen, and describled in \cite{anderson2022unified} when the WPGSD method is selected. The following example demonstrates a detailed implementation of adjusted-sequential p-values. The code to reproduce the example results can be found at \url{https://merck.github.io/wpgsd/articles/adj-seq-p.html}.

\begin{myexample}
\label{example: method -- weighted bonferroni -- nominal p}
  In a 2-arm controlled clinical trial example with one primary endpoint, there are 3 patient populations defined by the status of two biomarkers A and B: (i) the biomarker A positivepopulation, (ii) the biomarker B positive population, and (iii) the overall population. The 3 elementary hypotheses are: 
  \begin{itemize}
        \item $H_1$: experimental treatment is superior to control in the biomarker A positive population;
        \item $H_2$: experimental treatment is superior to control in the biomarker B positive population;
        \item $H_3$: experimental treatment is superior to control in the overall population.
    \end{itemize}
  The study has one interim analysis and one final analysis and the number of events is listed in Table \ref{table: example -- 3 pop -- event}.
  \begin{table}[htbp]
    \centering
    \caption{Number of events in Example 1A.
    \label{table: example -- 3 pop -- event}}
    \begin{adjustbox}{max width=0.95\textwidth}
    \begin{threeparttable}
    \begin{tabular}{ccc}
      \hline
      Population & Number of events at IA\textsuperscript{1} & Number of events at FA\textsuperscript{1}\\
      \hline
      Biomarker A positive & 100 & 200 \\
      Biomarker B positive & 110 & 220 \\
      Biomarker A B both positive & 80 & 160 \\
      Overall population & 225 & 450 \\
    \hline
    \end{tabular}
    \begin{tablenotes}
    \footnotesize
      \item[1] IA, FA is short for interim and final analysis, respectively.
    \end{tablenotes}
    \end{threeparttable}
    \end{adjustbox}
    \end{table}
  With the graphical approach review in Appendix \ref{appendix: graphic approach}, we assign the initial weights of $H_1, H_2, H_3$ as
  $
      \left(w_1(I), w_2(I), w_3(I) \right) 
      = 
      (0.3, 0.3, 0.4).
  $
  The multiplicity strategy is visualized in Figure \ref{fig: ex1 -- multi strategy}.
  We can compute the weighting strategy for each intersection hypothesis in Table \ref{table: example -- 3 pop -- weighting}. 
    If $H_1$ is rejected, then $3/7$ local significance level $\alpha_1$ will be propagated to $H_2$, and $4/7$ will go to $H_3$.
    If $H_3$ is rejected, then half of $\alpha_3$ goes to $H_1$, and half goes to $H_2$. Mathematically, its transition matrix is 
    $$
    G =
    \left(
    \begin{array}{ccc}
      0   &  3/7  & 4/7 \\
      3/7 &   0   & 4/7 \\
      1/2 &  1/2  & 0    \\
    \end{array}
    \right).
    $$
  \begin{figure}
      \centering
      \includegraphics[width=0.5\textwidth]{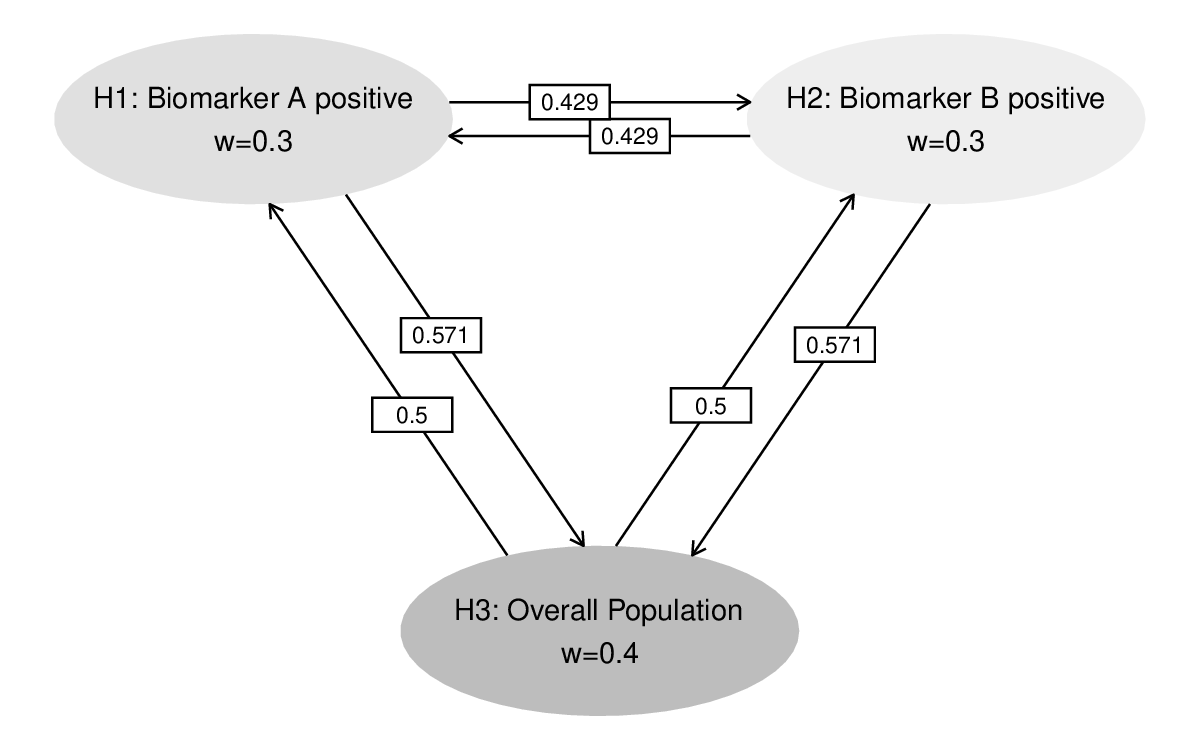}
      \caption{Graphical representation in Example 1A}
      \label{fig: ex1 -- multi strategy}
  \end{figure}
  Assume the nominal p-values of $H_1, H_2$, and $H_3$ at IA are 0.02, 0.01, 0.012 at the interim analysis, and 0.015, 0.012, 0.01 at the final analysis.
  For the weighted Bonferroni approach, each hypothesis employs an Hwang-Shib-DeCani (HSD; $\gamma = -4$) $\alpha$-spending function.
  For the WPGSD method, we assume an HSD($\gamma = -4$) $\alpha$-spending function is used for overall $\alpha$ spending with minimum information fraction (i.e., Method 3(b) in \cite{anderson2022unified}).
  Of note, the information fraction was 0.5 at the interim for all hypotheses given the number of events in Table \ref{table: example -- 3 pop -- event}.
  The correlation matrix of $\{Z_{j,k}\}_{j = 1, \ldots, m;\; k =  1, \ldots, K }$ is presented in Appendix \ref{appendix: correlated test stat} when using the WPGSD method. 
  With the methods above, we derive the adjusted-sequential p-values in Table \ref{table: example -- subpop -- weighted bonferroni seq-p}. 

  \begin{table}[htbp]
      \centering
      \caption{The p-values via Weighted Bonferroni in Example \ref{example: method -- weighted bonferroni -- nominal p}}
      \label{table: example -- subpop -- weighted bonferroni seq-p}
      \begin{adjustbox}{max width=0.95\textwidth}
      \begin{threeparttable}
      \begin{tabular}{c|cc|cc}
        \hline
        Hypothesis  
        & \multicolumn{2}{c}{Weighted Bonferroni} 
        & \multicolumn{2}{c}{WPGSD} \\
        \cline{2-5}
        & sequential p-values 
        & adjusted-sequential p-values 
        & sequential p-values 
        & adjusted-sequential p-values  \\
        \hline
        & \multicolumn{4}{c}{Interim analysis} \\
        \cline{2-5}
        $H_{1,2,3}$ & 0.2097  & - & 0.1636  & - \\
        $H_{1,2}$   & 0.1678  & - & 0.1400  & - \\
        $H_{1,3}$   & 0.1468  & - & 0.1302  & -\\
        $H_{2,3}$   & 0.1468  & - & 0.1282 & -  \\
        $H_{1}$     & 0.1258  & 0.2097 & 0.1258 & 0.1636 \\
        $H_{2}$     & 0.0839  & 0.2097 & 0.0839	& 0.1636 \\
        $H_{3}$     & 0.0839  &	0.2097 & 0.0839	& 0.1636 \\
        \hline
        & \multicolumn{4}{c}{Final analysis} \\
        \cline{2-5}
        $H_{1,2,3}$ & 0.0266 & - & 0.0206 & - \\
        $H_{1,2}$   & 0.0255 & - & 0.0210  & - \\
        $H_{1,3}$   & 0.0186 & - & 0.0165  & - \\
        $H_{2,3}$   & 0.0186 & - & 0.0162	  & - \\
        $H_{1}$     & 0.0159 &  0.0266 & 0.0159 & 0.0210 \\
        $H_{2}$     & 0.0127 &	0.0266 & 0.0127	& 0.0210 \\
        $H_{3}$     & 0.0106 &	0.0266 & 0.0106	&0.0206 \\
        \hline
      \end{tabular}
      \end{threeparttable}
      \end{adjustbox}
  \end{table}
  
  \textbf{Results by weighted Bonferroni method.}
  At the interim analysis, all the 3 adjusted-sequential p-values are greater than 0.025, so none are rejected. At the final analysis, all of them are larger than 0.025, and again no hypothesis is rejected. Besides the adjusted-sequential p-values, users can also apply the sequential p-values for individual and intersection hypotheses. Since consonance holds under weighted Bonferroni in group sequential designs \citep{maurer2013multiple, Hommel2007} there is no need for testing intersection hypotheses. 
  As indicated by Table \ref{table: example -- subpop -- weighted bonferroni seq-p}, at the final analysis, the sequential p-values for $H_1$ amount to 0.0266, which exceeds  $0.3 \times 0.025 = 0.0075$ where $0.3$ represents the initial weight assigned to $H_1$. As a result, we cannot reject $H_1$. Similarly, $H_2$ and $H_3$ cannot be rejected because $0.0266 > 0.3 \times 0.025 = 0.0075$ and $0.0266 \geq 0.4 \times 0.025 = 0.01$. 

  \textbf{Results by WPGSD.} Since the adjusted-sequential p-values at the interim analysis are all greater than 0.025, no hypotheses are rejected at that time. At the final analysis, all individual hypotheses are rejected since their adjusted-sequential p-values are all smaller than 0.025.
\end{myexample}

From the above example, we find the WPGSD method is more efficient than weighted Bonferroni since the sequential p-values for the intersection hypotheses are all smaller for WPGSD than weighted Bonferroni, which leads to more hypotheses being rejected. This is because the WPGSD accounts for the correlation between the
test statistics, which can increase study power or save sample size, compared to procedures not accounting for the correlation. For the consumer of trial results (e.g., the data monitoring committee), the adjusted sequential p-values are immediately interpretable to understand which hypotheses are rejected.

\section{Discussion}
\label{sec: discussion}
In multiple comparisons, there are two approaches to adjust for multiplicity. One approach is to \textbf{adjust the significance levels} to compare with nominal p-values observed. The other approach is to \textbf{adjust the p-values} and compare the adjusted p-values with the overall family-wise significance level for a trial (e.g., one-sided 0.025). 
In our previous work \citep{anderson2022unified}, we considered adjusted significance levels. In this paper, we proposed the adjusted sequential p-value as a simple-to-interpret method to describe the outcome of testing of individual hypotheses in group sequential designs. That is, the adjusted sequential p-value for any individual hypothesis need only be compared with the FWER for the combined set of hypotheses being tested. 
This is an extension of similar methods for fixed designs in 
\cite{xi2017unified}.  
From the computational perspective, the adjusted p-value approach is easier than the adjusted significance level approach for the weighted parametric method in fixed design (i.e., no interim analysis) \citep{xi2017unified}. However, this advantage cannot be extended to WPGSD. This is due to a need for interaction hypotheses since consonance for the WPGSD method which is not needed not ensured. For the weighted Bonferroni method, the consonance property holds and there is a short-cut as shown in Section \ref{sec: adj-seq p in wpgsd and bonferroni} with a sequential p-value approach.
While the computations for the adjusted sequential p-value are detailed, we have provided software in \url{https://github.com/Merck/wpgsd}. There are two alternatives to approach this testing problem:
\begin{enumerate}
\item The adjusted significance bound at each analysis for each hypothesis can be computed for all possible alpha levels at which each hypothesis can be tested in different intersection hypotheses \citep{anderson2022unified}. With, for example, 3 hypotheses, 3 analyses, and 3 possible alpha levels, this is already 9 bounds that would be required prior to unblinding for analysis in order to cover all possible needed evaluations. While the standard group sequential computations required are straightforward, interpretation can be challenging for a Data Monitoring Committee (DMC). However, this can be very useful for checking rejections made with the adjusted sequential p-value method we propose here.
\item	For weighted Bonferroni tests where no test correlations are accounted for, computing a sequential p-value for each hypothesis at each analysis can be a useful simplification. These sequential p-values can be plugged into a hypothesis graph as if no interim analyses were performed. Since the software is readily available for this in the \textit{gMCP} R package, this is straightforward. 
\end{enumerate}
In this paper, we have extended the sequential p-value approach in a group sequential design from an elementary hypothesis as in \cite{AdaptExtend} to intersection hypotheses and then to closed testing of individual hypotheses. Additionally, we provide the definition of the adjusted-sequential p-value for each elementary hypothesis. If the adjusted-sequential p-value of a hypothesis is less than the family-wise significance level, this hypothesis is rejected, whether at interim analyses or final analysis.
We apply the aforementioned adjusted inference in the weighted Bonferroni method and the WPGSD method accounting for test correlations, with a graphical approach. The attraction of WPGSD is increased testing efficiency by accounting for known correlations between individual tests. The adjusted p-value approach simplifies evaluation for a DMC and complements previous literature which mostly uses the adjusted significance level approach \citep{anderson2022unified}.

\bibliographystyle{apalike}
\bibliography{ccs}

\newpage
\section*{Appendix}
\label{sec: appendix}
\appendix

\section{Simplification of sequential p-value}
\label{appendix: seq-p simplify}

The sequential p-value of an elementary hypothesis $H_j$ at the $k$-th analysis can be simplified from \eqref{equ: def of seq p of H_j} as 
\begin{eqnarray*}
  p_{j,k}^{seq}
  & = & \nonumber
  \inf 
  \left\{ 
    \mu: 
    Z_{j,1} \geq  \widetilde Z_{j,1}(\mu)
    \; \cup \; 
    Z_{j,2} \geq  \widetilde Z_{j,2}(\mu)
    \;  \cup \; 
    \ldots
    \;  \cup \; 
    Z_{j,k} \geq  \widetilde Z_{j,k}(\mu)
  \right\} \\
  & = & 
  \sup
  \left\{
    \mu:
    \underset{1\leq k' \leq k}{\cap}
    \{Z_{j,k'} <  \widetilde Z_{j,k'}(\mu)\}
  \right\}\\
  & = &
  \nonumber
  \sup
  \left\{
    \mu:
    \underset{1\leq k' \leq k}{\max}
    \{Z_{j,k'} -  \widetilde Z_{j,k'}(\mu)\} < 0
  \right\}.
\end{eqnarray*}
With the above simplification, we can re-write the sequential p-values of an intersection hypothesis $H_J$ as
\begin{eqnarray*}
  p_{J,k}^{seq}
  & = & 
  \sup
  \left\{ 
    \mu: 
    \max_{1 \leq k' \leq k, \;  j \in J} 
    \{Z_{j,k'} -  \widetilde Z_{j, k'}(\mu, J)\} \leq 0 
  \right\},
\end{eqnarray*}
where $Z_{j,k'}$ is the test statistic for the elementary hypothesis $H_j$ at the $k'$-th analysis
and $\widetilde Z_{j, k'}(\mu, J)$ is the upper bound of the elementary hypothesis $H_j$ at the $k'$-th analysis when tested at level  $\mu$ in the intersection set $J$.
We note this requires a $|J|$-variate normal computations which are not demonstrated here but are available in the software repository at \url{https://github.com/Merck/wpgsd}.

\section{Graphical hypothesis testing  approach}
\label{appendix: graphic approach}

This section introduces the graphical method and its weighting strategy. It is independently derived by \cite{bretz2009graphical} and \cite{burman2009recycling} for the multiplicity problem. 
Though the graphical representations and rejection algorithms in these two articles are different, the underlying ideas are closely related \citep{guilbaud2011confidence}.
Using the graphical approach of \cite{bretz2009graphical}, the hypotheses $H_1, H_2, \ldots, H_m$ are represented by vertices with associated weights denoting the local significance levels $\alpha_1, \alpha_2, \ldots, \alpha_m$ with $\sum_{j=1}^m\alpha_j=\alpha,$ the FWER for all testing.
We further denote $I = \{1, 2, \ldots, m\}$ and $w_j=\alpha_j/\alpha, j= 1,2,\ldots,m$.
Any two vertices $H_i$ and $H_j, 1\le i,j\le m$ are connected through directed edges, where the associated weight $g_{ij}$ indicates the fraction of the (local) significance level $\alpha_i$ that is propagated to $H_j$ once $H_i$ (the hypothesis at the tail of the edge) has been rejected. 
A weight $g_{ij} = 0$ indicates that no propagation of the significance level is foreseen and the edge is dropped for convenience. 
Generally, for any $i, j \in I$, we have $0 \leq g_{ij} \leq 1$ and $\sum_{j=1}^m g_{ij} \leq 1$. 
We note $g_{ii} = 0, i=1,2,\ldots,k$. 
The matrix $G = (g_{ij}) \in \mathbb R^{m \times m}$ is referred to as the \textit{transition matrix}. 

For each $J \subseteq I$, there is a collection of weights $w_j(J)$ such that $0 \leq w_j(J) \leq 1$ and $\sum_{j \in J} w_j(J) \leq 1$.
Such $w_j(J)$ defines inference in the weighted Bonferroni test. Specifically, assume $p_j$ is the unadjusted p-value for $H_j$, then we reject $H_J$ if $p_j \leq w_j(J)\alpha$ for at least one $j\subseteq J$.
\cite{Hommel2007} introduced a useful subclass of sequentially rejective Bonferroni-based closed test procedures. They showed that the monotonicity condition
$$
  w_j(J_1) \leq w_j(J_2), \;\; \forall j \in J_2 \subseteq J_1 \subseteq J.
$$
ensures consonance. A closed procedure is called consonant if the rejection of the complete intersection null hypothesis $H_I$ further implies that at least one elementary hypothesis $H_j, (j \in I)$, can be rejected. 
This substantially simplifies the implementation and interpretation of related closed testing procedures, as the closure tree of $2^m - 1$ intersection hypotheses is tested in only $m$ steps. In practice, consonance is a desirable property leading to shortcut procedures that give the same rejection decisions as the original closed procedure but with fewer operations \citep{xi2017unified}.

From the above example, we find that, for a global null hypothesis $H_I$, its weighting strategies are formally defined by two components: (1) the weights of the elementary hypothesis is $w_i(I)$ for any $i \in I$, and (2) transition matrix $G$.
For an intersection hypothesis $H_J$ with $\emptyset \neq J \subseteq I$, the weights of each elementary hypothesis are $w_j(J)$ for any $j \in J$, which can be derived by Algorithm 1 in \cite{bretz2011graphical}.

\section{Weighting strategies of Example \ref{example: method -- weighted bonferroni -- nominal p} 
}

The weighting strategy of Example \ref{example: method -- weighted bonferroni -- nominal p} is shown in Table \ref{table: example -- 3 pop -- weighting}. 
  \begin{table}[htbp]
    \centering
    \caption{Weighting strategy in Example 1A and 1B.
    \label{table: example -- 3 pop -- weighting}}
    \begin{adjustbox}{max width=0.95\textwidth}
    \begin{threeparttable}
    \begin{tabular}{c|ccc}
      \hline
      $H_J$ & $w_1(J)$ & $w_2(J)$ & $w_3(J)$ \\
      \hline
      $H_{1,2,3}$ & 0.3 & 0.3 & 0.4 \\
      $H_{1,2}$ & 0.5 & 0.5 & - \\
      $H_{1,3}$ & 0.3 &  -  & 0.7 \\
      $H_{2,3}$ & -   & 0.3 & 0.7 \\
      $H_{1}$   & 1   & -   & - \\
      $H_{2}$   & -   & 1   & - \\
      $H_{3}$   & -   & -   & 1 \\
      \hline
    \end{tabular}
    \begin{tablenotes}
    \footnotesize
      \item[1] $H_{1,2,3}$ refers to the intersection hypothesis consisting $H_{1}, H_2$ and $H_3$, i.e., $H_1 \cap H_2 \cap H_3$. Similar logic also applies to $H_{1,2}, H_{1,3}$ and $H_{2,3}$.
    \end{tablenotes}
    \end{threeparttable}
    \end{adjustbox}
    \end{table}

\section{Correlated test statistics}
\label{appendix: correlated test stat}
As noted in \cite{chen2021multiplicity}, the test statistics $\{Z_{i,k}\}_{i \in I, k = 1, \ldots, K}$ are correlated via analysis time or overlapping populations or shared control arm:
$$
  \text{Corr}(Z_{i_1, k_1}, Z_{i_2, k_2})
  = 
  \frac{n_{i_1 \wedge i_2, k_1 \wedge k_2}}{\sqrt{n_{i_1,k_1} n_{i_2, k_2}}}.
$$
Here $n_{i,k}$ is the number of observations (or the number of events for time-to-event endpoints) collected cumulatively through analysis $k$ for elementary hypothesis $H_i$.
And $n_{i_1 \wedge i_2, k_1 \wedge k_2}$ is the number of observations (or events) included in both $Z_{i_1,k_1}$ and $Z_{i_2, k_2}$.
The full correlation matrix of $\{Z_{i,k}\}_{i \in I, k = 1, \ldots, K}$ ($mK \times mK$) can be derived this way and is referred to as the complete correlation structure (CCS) in \cite{chen2021multiplicity} and \cite{anderson2022unified}.
In Example \ref{example: method -- weighted bonferroni -- nominal p}, we present a numerical example and show the explicate calculation of the CCS.

\begin{myexample}
\label{example: method -- corr}
    
    Following Example \ref{example: method -- weighted bonferroni -- nominal p} with the number of events at the interim and the final analyses for each population in Table \ref{table: example -- 3 pop -- event}, we articulate the details of compute the correlation matrix as follows.
    
    In the interim analysis, the correlation between $H_1$ and $H_2$ is $\text{Corr}(Z_{1,1}, Z_{2, 1}) = \frac{64}{\sqrt{80 \times 88} }$. The numerator is the number of events included in both $Z_{1,1}, Z_{2, 1}$, and the denominator are the number of events of $Z_{1,1}, Z_{2, 1}$, respectively. 
    In the final analysis, the correlation between $H_2$ and $H_3$ is $\text{Corr}(Z_{2,2}, Z_{3, 2}) = \frac{176}{\sqrt{176 \times 360} }$. The numerator is the number of events included in both biomarker B positive and the overall population, and the denominator is the number of events of the two populations, respectively. 
    The full correlation matrix  in Table \ref{table: example -- 3 pop -- corr}
\end{myexample}

\begin{table}[htbp]
\centering
\caption{Correlation matrix of test statistics in Example \ref{example: method -- weighted bonferroni -- nominal p}.
\label{table: example -- 3 pop -- corr}}
\begin{adjustbox}{max width=0.95\textwidth}
\begin{threeparttable}
\begin{tabular}{c|cccccc}
  \hline
           & $Z_{1,1}$ & $Z_{2,1}$ & $Z_{3,1}$ & $Z_{1,2}$ & $Z_{2,2}$ & $Z_{3,2}$ \\
  \hline
  $Z_{1,1}$
  & 1 
  & $\frac{64}{\sqrt{80 \times 88}}$ 
  & $\frac{80}{\sqrt{80 \times 180}}$ 
  & $\frac{80}{\sqrt{80 \times 160}}$ 
  & $\frac{64}{\sqrt{80 \times 176}}$ 
  & $\frac{80}{\sqrt{80 \times 360}}$ \\
  $Z_{2,1}$
  & 0.76
  & 1
  & $\frac{88}{\sqrt{88 \times 180}}$ 
  & $\frac{64}{\sqrt{88 \times 160}}$ 
  & $\frac{88}{\sqrt{88 \times 176}}$  
  & $\frac{88}{\sqrt{88 \times 360}}$ \\
  $Z_{3,1}$
  & 0.67
  & 0.70 
  & 1
  & $\frac{80}{\sqrt{180 \times 160}}$
  & $\frac{88}{\sqrt{180 \times 176}}$
  & $\frac{180}{\sqrt{180 \times 360}}$ \\
  $Z_{1,2}$
  & 0.71
  & 0.54  
  & 0.47 
  & 1
  & $\frac{128}{\sqrt{160 \times 176}}$
  & $\frac{160}{\sqrt{160 \times 360}}$ \\
  $Z_{2,2}$
  & 0.54
  & 0.71  
  & 0.49 
  & 0.76
  & 1
  & $\frac{176}{\sqrt{176 \times 360}}$ \\
  $Z_{3,2}$
  & 0.47
  & 0.49 
  & 0.71
  & 0.67
  & 0.70
  & 1 \\
  \hline
\end{tabular}
\end{threeparttable}
\end{adjustbox}
\end{table}

We acknowledge that, in some cases, the correlation among the hypotheses is only known for subsets of the hypotheses. For example, suppose there are 4 elementary hypotheses in a trial: 
\begin{itemize}
    \item $H_1$: PFS in the biomarker-positive population;
    \item $H_2$: PFS in the overall population;
    \item $H_3$: OS in the biomarker-positive population;
    \item $H_4$: OS in the overall population.
\end{itemize}
Based on the overlapping number of events, the correlation is known between hypotheses concerning the same endpoint (i.e., $H_1$ and $H_2$ for PFS, and $H_3$ and $H_4$ for OS), but not between hypotheses concerning different endpoints (e.g., $H_1$ and $H_3$, or $H_1$ and $H_4$). When the correlation between test statistics is partially known, we can partition $I$ into $l$ mutually exclusive subsets $I_1, \ldots, I_l$ such that $I = \cup_{i=1}^{l} I_i$. For a test hypothesis $J \subseteq I$, let $J_i = J \cap I_i$ with $i = 1, \ldots, l$. Since the correlation of $J_i$ is fully known, we can apply the aforementioned method for inference within that subset of the intersection hypothesis.

\end{document}